\documentclass{ws-ijmpcs}

\usepackage{color}

\begin{document}

\markboth{C.-M. Chen and J.-R. Sun} {The Kerr-Newman/CFTs
correspondence}

%
\catchline{}{}{}{}{}
%

\title{The Kerr-Newman/CFTs Correspondence}

\author{Chiang-Mei Chen}

\address{Department of Physics and Center for Mathematics and Theoretical Physics,\\
National Central University,\\ Chungli 320, Taiwan\\
cmchen@phy.ncu.edu.tw}

\author{Jia-Rui Sun}

\address{Department of Physics, National Central University,\\
Chungli 320, Taiwan\\
jrsun@phy.ncu.edu.tw}

\maketitle

\begin{history}
\received{Day Month Year}
\revised{Day Month Year}
\end{history}

\begin{abstract}
In this article, we review recent studies on multiple dual 2D CFT
descriptions of the Kerr-Newman black hole, in terms of the Kerr/CFT
and Reissner-Nordstr\"om/CFT correspondences. A microscopic hair
conjecture is suggested.

\end{abstract}

\ccode{PACS numbers: }

\section{Introduction}
The near horizon geometries of certain extremal black holes contain
the AdS$_2$ or AdS$_3$ structures, which are known to play important
role in understanding the macroscopic and microscopic entropies of
black holes long ago. The well known examples include the
Bertotti-Robinson solution (AdS$_2$) for extremal
Reissner-Nordstr\"om (RN) black hole in Einstein-Maxwell
theory~\cite{Bertotti:1959pf,Robinson:1959ev}, and the near horizon
geometry (AdS$_3$) of the D1-D5 system in string
theory~\cite{Strominger:1996sh}. Based on the holographic
principle~\cite{'tHooft:1993gx,Susskind:1994vu} and its realization
in the asymptotically AdS spacetime, one can relate the quantum
gravity description of black holes with their dual CFT in terms of
the AdS/CFT
correspondence~\cite{Maldacena:1997re,Gubser:1998bc,Witten:1998qj}.
In the past three years, new advances have been made in studying the
CFT descriptions of a class of extremal black holes in which the
near horizon geometries are of the warped AdS$_3$ structure (with
$SL(2, R)_R \times U(1)_L$ isometry). The initial progress was the
calculation of the left hand central charge and temperature of the
2D CFT dual to the extremal Kerr black hole, called the Kerr/CFT
correspondence~\cite{Guica:2008mu}.
The Kerr/CFT duality was further generalized into the near extremal
and nonextremal cases by comparing the scattering amplitudes of
external probe fields from the gravity side with those of the
corresponding operators in the dual 2D CFT
side~\cite{Bredberg:2009pv,Castro:2010fd}. Especially, even though
the nonextremal Kerr black hole doesn't contain the warped AdS$_3$
geometry, a 2D hidden conformal symmetry can be detected by a probe
scalar field at low frequencies. Other nonextremal black holes are
shown to have the similar property,
see~\cite{Chen:2010as,Wang:2010qv,Chen:2010xu,Becker:2010dm,Chen:2010bh,Chen:2010yu,Chen:2010ywa}.
There are various related studies and
generalizations~\cite{Matsuo:2009sj,Castro:2009jf,Cvetic:2009jn,Hartman:2008pb,Garousi:2009zx,Chen:2010bs,Chen:2009ht,Lu:2008jk,Chow:2008dp,Isono:2008kx,Wu:2009di,Chen:2009xja,Chen:2010ni,Becker:2010jj},
see a recent review~\cite{Bredberg:2011hp} for more references
therein.

A nontrivial generalization of the Kerr/CFT correspondence is the
studying of the CFT descriptions for nonrotating charged black
holes, the simplest example is the (near) extremal 4D RN black hole,
which only contains a near horizon AdS$_2$ geometry, but accompanied
by an $U(1)$ gauge field. The key observation was that the
background $U(1)$ gauge field plays an equivalent role as that of
the rotation in rotating black holes, consequently a warped AdS$_3$
structure is formed together with the AdS$_2$ base manifold. This
fact enables us to investigate the (near) extremal RN/CFT
duality~\cite{Hartman:2008pb,Garousi:2009zx,Chen:2010bs,Chen:2009ht},
and the 2D CFT description for nonextremal RN black
hole~\cite{Chen:2010as,Chen:2010yu}.

Combined with the results in the Kerr/CFT and RN/CFT
correspondences, it is expectable that a multiple dual 2D CFT
descriptions should exist for charged rotating black holes, e.g.,
the Kerr-Newman (KN) black hole. We showed that, besides the angular
momentum $J$-picture (associated with the Kerr/CFT duality, where
the central charges of the dual 2D CFT are $c_L = c_R = 12J$)
revealed in~\cite{Hartman:2008pb,Wang:2010qv,Chen:2010bh}, another
$Q$-picture (associated with the RN/CFT duality, in which the
central charges are $c_L = c_R = 6Q^3/\ell$, where $\ell$ is the
parameter depends on the embedding) does exist for the general
nonextremal KN black hole by probing the 2D hidden conformal
symmetries via an external charged scalar field~\cite{Chen:2010ywa}.
Accordingly, a KN/CFTs correspondence was suggested. This suggestion
was checked by matching of the absorption cross sections and real
time correlators calculated in both the $J$- and the $Q$-pictures
from the gravity and the CFT sides. It is natural to believe that
these two individual descriptions are just special limits of a
unknown full holographic dual description of the KN black hole. The
KN/CFTs duality could be much more easily understood in the
geometric picture since in the near extremal limit, the near horizon
KN black hole contains an AdS$_2$ geometry together with two $U(1)$
fibers, with $SL(2, R)_R \times U(1) \times U(1)$ isometry. Thus
there are two possible choices to form the desired warped AdS$_3$
structures, and certainly to have two individual dual 2D CFT
descriptions. A similar idea has been studied later for the spinning
M5 branes in~\cite{Compere:2010uk}. Moreover, our work also
suggested a fascinating ``microscopic hair conjecture''
--- for each macroscopic hair parameter, in additional to the mass of a
black hole in the Einstein-Maxwell theory (with dimension $D\geq
3$), there should exist an associated holographic CFT$_2$
description.

This article is organized as follows. We first review some basic
properties of the KN black hole and study the scattering of a probe
charged scalar field in section II. In
sections III and IV, we explicitly analyze the $J$-picture and the
$Q$-picture of the KN black hole, respectively, including the
probing of hidden conformal symmetry, the calculation of absorption
cross sections and real time correlators. Then we give the
conclusion and discussion in section V.

\section{Charged Scalar Field in the Kerr-Newman Background}
The 4D Kerr-Newman black hole in the Boyer-Lindquist coordinates is
\begin{eqnarray}
ds^2 &=& - \frac{\Delta - a^2 \sin^2\theta}{\Sigma} \left[ dt +
\frac{(2 M r - Q^2) a \sin^2\theta}{\Delta - a^2 \sin^2\theta}
d\phi \right]^2
\nonumber\\
&& + \Sigma \left( \frac{dr^2}{\Delta} + d\theta^2 + \frac{\Delta \sin^2\theta}{\Delta - a^2\sin^2\theta}
d\phi^2 \right),
\nonumber\\
A_{[1]} &=& - \frac{Q r}{\Sigma} \left( dt - a \sin^2\theta d\phi \right),
\end{eqnarray}
where the three macroscopic hairs are the mass $M$, electric charge
$Q$ and angular momentum $J =  Ma$, and
\begin{equation}
\Sigma = r^2 + a^2 \cos^2\theta, \qquad \Delta = r^2 - 2 M r + a^2 + Q^2.
\end{equation}
The radii of black hole outer and inner horizons $r_\pm$, the
horizon angular velocity $\Omega_H$ and the chemical potential
$\Phi_H$ are
\begin{equation}
r_\pm = M \pm \sqrt{M^2 - a^2 - Q^2}, \qquad \Omega_H =
\frac{a}{r_+^2 + a^2}, \qquad \Phi_H = \frac{Q r_+}{r_+^2 + a^2}.
\end{equation}
In addition, the Hawking temperature and the black hole entropy, are
\begin{equation}
T_H = \frac{\kappa}{2 \pi} = \frac{r_+ - r_-}{4\pi (r_+^2 + a^2)},
\qquad S_{BH} = \frac{A_+}{4} = \pi (r_+^2 + a^2),
\end{equation}
where $\kappa$ and $A_+$ are the surface gravity and area of the
outer horizon, respectively.

The Klein-Gordon (KG) equation of a massive probe charged scalar
field $\Phi$ scattering in the KN black hole is
\begin{equation}
(\nabla_{\alpha} - i q A_{\alpha})(\nabla^{\alpha} - i q A^{\alpha})
\Phi = \mu^2\Phi,
\end{equation}
where $\mu$ and $q$ are mass and electric charge parameters of the
scalar field respectively. After assuming the following mode
expansions
\begin{equation}
\Phi(t, r, \theta, \phi) = \mathrm{e}^{- i \omega t + i m \phi} R(r)
S(\theta),
\end{equation}
the KG equation can be decoupled into the angular and radial
equations as
\begin{eqnarray}
\frac1{\sin\theta} \partial_\theta (\sin\theta \, \partial_\theta S) - \left[ a^2 (\omega^2 - \mu^2) \sin^2\theta +
\frac{m^2}{\sin^2\theta} - \lambda \right] S &=& 0,
\\
\partial_r (\Delta \partial_r R) + \left[ \frac{[ (r^2 + a^2) \omega - q Q r - m a ]^2}{\Delta} - \mu^2 (r^2 + a^2) + 2 m a \omega - \lambda \right] R &=& 0,
\end{eqnarray}
where $\lambda$ is the separation constant. When $\mu = 0$, the
radial equation can be further expressed in a symmetric form
\begin{eqnarray}\label{radial}
&& \partial_r (\Delta \partial_r R) + \left[ \frac{\left[ (r_+^2 +
a^2) \omega - a m - Q r_+ q \right]^2}{(r - r_+) (r_+ - r_-)} -
\frac{\left[ (r_-^2 + a^2) \omega - a m - Q r_- q \right]^2}{(r -
r_-) (r_+ - r_-)} \right] R
\nonumber\\
&& \quad + \left[ \omega^2 r^2 + 2 (\omega M - q Q) \omega r +  \omega^2
a^2 - \omega^2 Q^2 + (2 \omega M - q Q)^2 \right] R = \lambda R.
\end{eqnarray}
Under conditions: (1) small frequency $\omega M \ll 1$ (consequently
$\omega a \ll 1$ and $\omega Q \ll 1$), (2) small probe charge $q Q
\ll 1$ and (3) near region $\omega r \ll 1$, the potential terms in
the second line can be neglected, and $\lambda=\lambda_l= l (l +
1)$. Then eq.(\ref{radial}) reduces to
\begin{equation} \label{eqR}
\partial_r (\Delta \partial_r R) \!+\! \left[ \frac{\left[ (r_+^2 + a^2) \omega
- a m - Q r_+ q \right]^2}{(r - r_+) (r_+ - r_-)} \!-\! \frac{\left[
(r_-^2 + a^2) \omega - a m - Q r_- q \right]^2}{(r - r_-) (r_+ -
r_-)} \right] R \!=\! \lambda_l R.
\end{equation}

In the following two sections, we will show how to probe the twofold
2D conformal symmetries hidden in eq.(\ref{eqR}). We first reproduce
the $J$-picture in terms of the Kerr/CFT correspondence studied
in~\cite{Wang:2010qv,Chen:2010xu}, and then investigate the new
$Q$-picture in terms of the RN/CFT correspondence. We just list the
main results, for detailed calculation, please refer
to~\cite{Chen:2010ywa}.

\section{Angular Momentum ($J$-) Picture}

\subsection{Hidden Conformal Symmetry}
To probe the $J$-picture CFT$_2$ description dual to the KN black
hole, one should consider a neutral scalar field by imposing $q =
0$, then eq.(\ref{eqR}) becomes
\begin{equation}
\left( \partial_r (\Delta \partial_r) - \frac{\left[ (r_+^2 + a^2)
\partial_t + a \partial_\phi \right]^2}{(r - r_+) (r_+ - r_-)} +
\frac{\left[ (r_-^2 + a^2) \partial_t + a \partial_\phi
\right]^2}{(r - r_-) (r_+ - r_-)} \right) \Phi = l (l + 1) \Phi.
\end{equation}
It was shown that the left hand side of the above equation is just
the $SL(2,R)$ Casimir operator acting on $\Phi$ with the
identifications of parameters~\cite{Chen:2010ywa}
\begin{equation}
T^J_L = \frac{r_+^2 + r_-^2 + 2 a^2}{4 \pi a (r_+ + r_-)}, \quad
T^J_R = \frac{r_+ - r_-}{4 \pi a}, \qquad n^J_L = - \frac1{2 (r_+ +
r_-)}, \quad n^J_R = 0.
\end{equation}
Since the neutral probe scalar field does not couple with the
background gauge field, it only reveals the angular momentum
dominated section of the dual field theory, resembling the 2D CFT
dual to the Kerr black hole with central charges
\begin{equation}
c^J_L = c^J_R = 12 J = 12 M a.
\end{equation}
One can directly check that the CFT microscopic entropy from the
Cardy formula indeed reproduces the Bekenstein-Hawking area entropy
of the KN black hole:
\begin{equation}
S^J_\mathrm{CFT} = \frac{\pi^2}3 \left( c^J_L T^J_L + c^J_R T^J_R
\right) = \pi (r_+^2 + a^2) = S_\mathrm{BH}.
\end{equation}

\subsection{Scattering}
The near region solutions of eq.(\ref{eqR}) include ingoing and
outgoing modes as
\begin{eqnarray}
R_J^\mathrm{(in)} &=& z^{- i \gamma_J} (1 - z)^{l + 1} \, F( a_J,
b_J; c_J; z ),
\nonumber\\
R_J^\mathrm{(out)} &=& z^{i \gamma_J} (1 - z)^{l + 1} \, F( a_J^*,
b_J^*; c_J^*; z ),
\end{eqnarray}
where $q = 0$ is imposed, $z = \frac{r - r_+}{r - r_-}$, the
expressions of $a_J$, $b_J$, $c_J$ and $\gamma_J$ are
\begin{eqnarray}
&& \gamma_J = \frac{\omega (r_+^2 + a^2) - m a}{r_+ - r_-}, \qquad a_J = 1 + l - i \frac{\omega (r_+^2 + r_-^2 + 2 a^2) - 2 m a}{r_+ - r_-},
\nonumber\\
&& b_J = 1 + l - i \omega (r_+ + r_-), \qquad c_J = 1 - i 2 \gamma_J.
\end{eqnarray}

The asymptotic form of the ingoing mode is
\begin{equation}\label{asymRJ}
R_J^\mathrm{(in)}(r \gg M) \sim A_J \, r^l + B_J \, r^{- l - 1},
\end{equation}
where
\begin{equation}
A_J = \frac{\Gamma(c_J) \Gamma(2 l + 1)}{\Gamma(a_J) \Gamma(b_J)},
\qquad B_J = \frac{\Gamma(c_J) \Gamma(- 2 l - 1)}{\Gamma(c_J - a_J)
\Gamma(c_J - b_J)},
\end{equation}
and the conformal weights of the operator dual to the scalar field
are
\begin{equation}
h^J_L = h^J_R = l + 1.
\end{equation}
Hence, $a_J$ and $b_J$ can be expressed in terms of $\omega^J_L$ and
$\omega^J_R$
\begin{equation}
a_J = h^J_R - i \frac{\omega^J_R}{2 \pi T^J_R}, \qquad b_J = h^J_L -
i \frac{\omega^J_L}{2 \pi T^J_L},
\end{equation}
with
\begin{equation}
\omega^J_L = \frac{\omega (r_+^2 + r_-^2 + 2 a^2)}{2 a}, \quad \omega^J_R = \frac{\omega (r_+^2 + r_-^2 + 2 a^2) - 2 m a}{2 a}.
\end{equation}
The essential part of the absorption cross section can be estimated as
\begin{equation}
P^J_\mathrm{abs} \sim |A_J|^{-2} \sim \sinh( 2 \pi \gamma_J) \,
\left| \Gamma(a_J) \right|^2 \, \left| \Gamma(b_J) \right|^2.
\end{equation}
Further with the help of the first law of thermodynamics,
\begin{equation}
\delta S^J_{CFT} = \frac{\delta E^J_L}{T^J_L} + \frac{\delta
E^J_R}{T^J_R}=\delta S_{BH} = \frac1{T_H} \delta M -
\frac{\Phi_H}{T_H} \delta Q - \frac{\Omega_H}{T_H} \delta J,
\end{equation}
the absorption cross section can be expressed as~\cite{Chen:2010xu}
\begin{eqnarray}
P^J_\mathrm{abs} &\sim& (T^J_L)^{2 h^J_L - 1} (T^J_R)^{2 h^J_R -
1} \sinh\left( \frac{\omega^J_L}{2 T^J_L} + \frac{\omega^J_R}{2
T^J_R} \right)
\nonumber\\
&& \times\left| \Gamma\left( h^J_L + i
\frac{\omega^J_L}{2 \pi T^J_L} \right) \right|^2 \, \left|
\Gamma\left( h^J_R + i \frac{\omega^J_R}{2 \pi T^J_R} \right)
\right|^2,
\end{eqnarray}
which is the finite temperature absorption cross section of an
operator dual to the neutral probe scalar field in the $J$-picture
2D CFT dual to the KN black hole.

\subsection{Real-time Correlator}
The two-point retarded correlator calculated from the gravity side
is~\cite{Son:2002sd,Chen:2010ni}
\begin{eqnarray}\label{GreenRJ}
G^J_R \sim \frac{B_J}{A_J} &=& (-)^{h^J_L + h^J_R} \sin\left(i \frac{\omega^J_L}{2 T^J_L} \right) \sin\left(i \frac{\omega^J_R}{2 T^J_R} \right)
\nonumber\\
&& \times \Gamma\left( h^J_L - i \frac{\omega^J_L}{2 \pi T^J_L} \right) \Gamma\left( h^J_L + i \frac{\omega^J_L}{2 \pi T^J_L} \right)
\nonumber\\
&& \times \Gamma\left( h^J_R - i \frac{\omega^J_R}{2 \pi T^J_R} \right) \Gamma\left( h^J_R + i \frac{\omega^J_R}{2 \pi T^J_R} \right).
\end{eqnarray}

From the CFT side, the retarded Green function $G_R(\omega_L,
\omega_R)$ can be obtained from the Euclidean correlator (in terms
of the Euclidean frequencies $\omega_{EL} = i \omega_L$, and
$\omega_{ER} = i \omega_R$)
\begin{eqnarray}\label{CFTGR}
G_E(\omega_{EL}, \omega_{ER}) &\sim& T_L^{2h_L -1} T_R^{2h_R - 1} \mathrm{e}^{i \frac{\tilde\omega_{EL}}{2 T_L}} \mathrm{e}^{i \frac{\tilde\omega_{ER}}{2 T_R}}
\nonumber\\
&& \times \Gamma\left( h_L - \frac{\tilde\omega_{EL}}{2 \pi T_L} \right) \Gamma\left( h_L + \frac{\tilde\omega_{EL}}{2 \pi T_L} \right)
\nonumber\\
&& \times \Gamma\left( h_R - \frac{\tilde\omega_{ER}}{2 \pi T_R} \right) \Gamma\left( h_R + \frac{\tilde\omega_{ER}}{2 \pi T_R} \right),
\end{eqnarray}
(where $\tilde\omega_{EL} = \omega_{EL} - i q_L \mu_L$ and
$\tilde\omega_{ER} = \omega_{ER} - i q_R \mu_R$) by the analytic
continuation from the on the upper half complex $\omega_{L,R}$-plane
\begin{equation}\label{CFTGER}
G_E(\omega_{EL}, \omega_{ER}) = G_R(i \omega_L, i \omega_R), \qquad
\omega_{EL}, \omega_{ER} > 0,
\end{equation}
and the Euclidean frequencies $\omega_{EL}$ and $\omega_{ER}$ should
take discrete values of the Matsubara frequencies ($m_L, m_R$ are
integers for bosons and half integers for fermions)
\begin{equation}\label{Matsubara}
\omega_{EL} = 2 \pi m_L T_L, \qquad \omega_{ER} = 2 \pi m_R T_R.
\end{equation}
At these frequencies, the retarded Green function precisely agrees
with the gravity side computation eq.(\ref{GreenRJ}) up to a
numerical normalization factor.

\section{Charge ($Q$-) Picture}

\subsection{Hidden Conformal Symmetry}
To probe the $Q$-picture CFT$_2$ description one should impose $m =
0$, and the the corresponding radial equation, from eq.(\ref{eqR}),
becomes
\begin{equation}
\left( \partial_r (\Delta \partial_r) - \frac{\left[ (r_+^2 +
a^2)
\partial_t + \frac{Q r_+}{\ell} \partial_\chi \right]^2}{(r - r_+) (r_+ -
r_-)} + \frac{\left[ (r_-^2 + a^2) \partial_t + \frac{Q r_-}{\ell}
\partial_\chi \right]^2}{(r - r_-) (r_+ - r_-)} \right) \Phi = l(l+1) \Phi,
\end{equation}
where $\partial_\chi$ acts on ``internal space'' of the $U(1)$
symmetry such that $\partial_\chi \Phi = i \ell q \Phi$, and $\ell$
is a parameter depending on the embedding,
see~\cite{Chen:2010bs,Chen:2010yu} for detailed explanation.
Similarly, the radial equation is just the Casimir operator of the
$SL(2,R)$ Lie algebra~\cite{Chen:2010ywa} acting on $\Phi$ with the
identifications
\begin{eqnarray}
T^Q_L = \frac{(r_+^2 + r_-^2 + 2 a^2) \ell}{4 \pi Q (r_+ r_- -
a^2)}, &\quad& T^Q_R = \frac{(r_+^2 - r_-^2) \ell}{4 \pi Q (r_+ r_-
- a^2)},
\\
n^Q_L = - \frac{r_+ + r_-}{4 (r_+ r_- - a^2)}, &\quad& n^Q_R = -
\frac{r_+ - r_-}{4 (r_+ r_- - a^2)}.
\end{eqnarray}
Now since the momentum mode on the $\phi$ direction is turned off,
such probe scalar field is not able to explore the information of
the background rotation. Therefore, it can reveal only the charge
dominated subsection resembling the 2D CFT dual to the RN black hole
with
\begin{equation}
c^Q_L = c^Q_R = \frac{6 Q^3}{\ell},
\end{equation}
and the CFT microscopic entropy matches the Bekenstein-Hawking area
entropy of the KN black hole
\begin{equation}
S^Q_\mathrm{CFT} = \frac{\pi^2}3 \left( c^Q_L T^Q_L + c^Q_R T^Q_R
\right) = \pi (r_+^2 + a^2) = S_\mathrm{BH}.
\end{equation}

\subsection{Scattering}
The solutions of the near region KG eq.(\ref{eqR}), with $m = 0$,
are
\begin{eqnarray}
R_Q^\mathrm{(in)} &=& z^{- i \gamma_Q}  (1 - z)^{l + 1} \, F( a_Q,
b_Q; c_Q; z ),
\nonumber\\
R_Q^\mathrm{(out)} &=& z^{i \gamma_Q}  (1 - z)^{l + 1} \, F( a_Q^*,
b_Q^*; c_Q^*; z ),
\end{eqnarray}
where the expressions of $a_Q$, $b_Q$, $c_Q$ and $\gamma_Q$ are
\begin{eqnarray}
&& \gamma_Q = \frac{\omega (r_+^2 + a^2) - q Q r_+}{r_+ - r_-}, \quad a_Q = 1 + l - i \frac{\omega (r_+^2 + r_-^2 + 2 a^2) - q Q (r_+ + r_-)}{r_+ - r_-},
\nonumber\\
&& b_Q = 1 + l - i \left[ \omega (r_+ + r_-) - q Q \right], \quad c_Q = 1 - i 2 \gamma_Q.
\end{eqnarray}

The asymptotic solution of the ingoing mode is
\begin{equation} \label{asymRQ}
R_Q^\mathrm{(in)}(r \gg M) \sim A_Q \, r^l + B_Q \, r^{- l - 1},
\end{equation}
where
\begin{equation}
A_Q = \frac{\Gamma(c_Q) \Gamma(2 l + 1)}{\Gamma(a_Q) \Gamma(b_Q)},
\qquad B_Q = \frac{\Gamma(c_Q) \Gamma(- 2 l - 1)}{\Gamma(c_Q - a_Q)
\Gamma(c_Q - b_Q)},
\end{equation}
and
\begin{equation}
h^Q_L = h^Q_R = l + 1.
\end{equation}
The coefficients $a_Q$ and $b_Q$ can be expressed as
\begin{equation}
a_Q = h^Q_R - i \frac{\tilde\omega^Q_R}{2 \pi T^Q_R}, \qquad b_Q =
h^Q_L - i \frac{\tilde\omega^Q_L}{2 \pi T^Q_L},
\end{equation}
where
\begin{equation}
\tilde\omega^Q_L = \omega^Q_L - q^Q_L \mu^Q_L, \qquad
\tilde\omega^Q_R = \omega^Q_R - q^Q_R \mu^Q_R,
\end{equation}
with
\begin{eqnarray}
&& \omega^Q_L = \frac{\ell \omega (r_+ + r_-)(r_+^2 + r_-^2 +
2a^2)}{2 Q (r_+ r_- - a^2)}, \quad q^Q_L = q, \quad \mu^Q_L =
\frac{\ell (r_+^2 + r_-^2 + 2a^2)}{2 (r_+ r_- - a^2)},
\nonumber\\
&& \omega^Q_R = \frac{\ell \omega (r_+ + r_-)(r_+^2 + r_-^2 +
2a^2)}{2 Q (r_+ r_- - a^2)}, \quad q^Q_R = q, \quad \mu^Q_R =
\frac{\ell (r_+ + r_-)^2}{2 (r_+ r_- - a^2)}.
\end{eqnarray}
The absorption cross section can be estimated by
\begin{equation}
P^Q_\mathrm{abs} \sim |A_Q|^{-2} \sim \sinh( 2 \pi \gamma_Q) ) \,
\left| \Gamma(a_Q) \right|^2 \, \left| \Gamma(b_Q) \right|^2.
\end{equation}
Again, with the help of the first law of thermodynamics $\delta
S^Q_{CFT} = \delta S_{BH}$, the absorption cross section can be
expressed as
\begin{eqnarray}
P^Q_\mathrm{abs} &\sim& (T^Q_L)^{2 h^Q_L - 1} (T^Q_R)^{2 h^Q_R - 1} \sinh\left( \frac{\tilde\omega^Q_L}{2 T^Q_L} +
\frac{\tilde\omega^Q_R}{2 T^Q_R} \right)
\nonumber\\
&& \times \left| \Gamma\left( h^Q_L + i \frac{\tilde\omega^Q_L}{2 \pi T^Q_L} \right) \right|^2 \, \left| \Gamma\left( h^Q_R + i \frac{\tilde\omega^Q_R}{2 \pi T^Q_R} \right) \right|^2,
\end{eqnarray}
which matches the finite temperature absorption cross section of an
operator with the conformal weights ($h_L, h_R$), frequencies
($\omega_L, \omega_R$) electric charges ($q_L, q_R$) and chemical
potentials ($\mu_L, \mu_R$) in the dual 2D CFT with the temperatures
($T_L, T_R$).

\subsection{Real-time Correlator}
In the $Q$-picture, the two-point retarded correlator is
\begin{eqnarray}
G^Q_R \sim \frac{B_Q}{A_Q} &=& (-)^{h^Q_L + h^Q_R} \sin\left(i \frac{\tilde\omega^Q_L}{2 T^Q_L} \right) \sin\left(i
\frac{\tilde\omega^Q_R}{2 T^Q_R} \right)
\nonumber\\
&& \times \Gamma\left( h^Q_L - i \frac{\tilde\omega^Q_L}{2 \pi T^Q_L} \right) \Gamma\left( h^Q_L + i \frac{\tilde\omega^Q_L}{2 \pi T^Q_L} \right)
\nonumber\\
&& \times \Gamma\left( h^Q_R - i \frac{\tilde\omega^Q_R}{2 \pi T^Q_R} \right) \Gamma\left( h^Q_R + i \frac{\tilde\omega^Q_R}{2 \pi T^Q_R} \right).
\end{eqnarray}
At the Matsubara frequencies~(\ref{Matsubara}), the retarded Green
function agrees well with the CFT results~(\ref{CFTGR}-\ref{CFTGER})
up to a normalization factor depending on $q^Q_L$ and $q^Q_R$.

\section{Conclusion and Discussion}
In this short article, we briefly reviewed our recent work on the
twofold dual 2D CFT descriptions of the generic nonextremal KN black
hole by probing the hidden conformal symmetries, in terms of the
Kerr/CFT and RN/CFT correspondences. The KN/CFTs dualities are shown
to be supported by the matching of the entropies, absorption cross
sections and real time correlators computed from both the gravity
and the CFT sides.


A geometric way to understand this multiple dual CFT pictures is
that, the KN black hole contains two $U(1)$ fibers, one is the
rotation another is the electromagnetic field. Its near extremal
near horizon geometry has the isometry $SL(2,R)_R\times U(1)\times
U(1)$, thus there are two options to form the desired warped AdS$_3$
structures, consequently, leading two dual 2D CFT dualities.
Therefore, besides the mass, each
one of the other two macroscopic hairs of the KN black hole, can
provide an individual holographically dual CFT$_2$ description.
This multiple dual CFT descriptions has been discussed for the higher
dimensional rotating black holes with multiple rotations, in which
each angular momentum can provide an individual dual CFT$_2$, see
for example~\cite{Lu:2008jk,Chen:2009xja}. It would be interesting
to check our microscopic hair conjecture for more general black
holes with multiple $U(1)$ charges, and find a full dual field
theory description in which each CFT corresponds to some certain
limit.

\section*{Acknowledgement}
This work was supported by the National Science Council of the R.O.C. under the grant NSC 99-2112-M-008-005-MY3 and in part by the National Center of Theoretical Sciences (NCTS).


\end{document}